\documentclass[prl,twocolumn,aps,preprintnumbers,floatfix,showpacs]{revtex4}
\usepackage[dvips]{graphicx}
\usepackage{latexsym}
\usepackage{revsymb}
\usepackage{amsfonts}
\usepackage{amsmath}
\usepackage{amssymb}
\usepackage{bm}

\begin{document}

\preprint{VT-IPNAS-10-05}

\title{Quantum Gravity, Dynamical Energy-Momentum Space and Vacuum Energy}
\author{Lay Nam Chang}\email{laynam@vt.edu}
\author{Djordje Minic}\email{dminic@vt.edu}
\author{Tatsu Takeuchi}\email{takeuchi@vt.edu}
\affiliation{Institute for Particle, Nuclear and Astronomical Sciences, Department of Physics, Virginia Tech, Blacksburg, VA 24061, U.S.A.} 

\date{\today}

\begin{abstract}
We argue that the combination of the principles of quantum theory and general relativity
allow for a dynamical energy-momentum space.
We discuss the freezing of vacuum energy in such a dynamical energy-momentum space and present 
a phenomenologically viable seesaw formula for the cosmological constant in this context.
\end{abstract}


\maketitle


The origins of the cosmological constant remain shrouded in mystery and represent a central issue in theoretical physics \cite{reviews}.  The question has assumed added urgency because of recent observations that it is small, positive, and clearly non-zero \cite{cc}.

The cosmological constant problem seems to violate our prejudice about
decoupling of short distance (UV) and long distance (IR) scales,
which underlies the use of effective field theory.
The cosmological constant $\Lambda$, being interpretable both as
the vacuum energy and as the scale of the observed universe, 
goes against this notion and suggests a mixing between UV and IR physics.

More specifically, the cosmological constant is
the leading operator in the effective action
describing the Standard Model (SM) coupled to gravity 
$S_i = \Lambda \int\!\sqrt{g},$
where on one side $\Lambda$ is tied to the Hubble scale $h$ as $\Lambda \sim h^{-2}$,
and on the other, it is directly related to the properly regulated sum of the vacuum modes 
$\int d^3\mathbf{p}\, \frac{1}{2} \hbar \omega_{p}$.
The vacuum energy is also naturally interpretable as the minimum of 
the effective potential associated with the particle physics sector. 
The UV/IR mixing feature of the cosmological constant can be summarized by 
a numerological formula
for the observed vacuum energy, i.e. the cosmological
constant seesaw formula
\begin{equation}
\epsilon \;\sim\; \frac{M_\mathrm{SM}^2}{M_P}
\label{ccseesaw}
\end{equation}
involving the vacuum energy scale $\epsilon$, the particle physics scale
$M_\mathrm{SM}$ ($\sim 1~\mathrm{TeV}$), and the Planck scale $M_P$.
According to this formula, the dimension of the cosmological constant operator 
is drastically anomalous \cite{banks}.

This formula is reminiscent of the well-known
seesaw mechanism used to explain the smallness of neutrino masses \cite{neutrino}.
In this letter,
we attempt to provide a framework for the putative cosmological constant seesaw mechanism 
that conceptually transcends the usual one used in effective field theories.

What could be the basic physics behind  
UV/IR mixing and the large effective anomalous
dimension for the vacuum energy operator as indicated by 
Eq.~(\ref{ccseesaw})?
One way to obtain the required scaling is by 
changing the momentum space measure in the sum over vacuum energy modes \cite{planck}
\begin{equation}
\int^{\mu} d^3\mathbf{p}\, \frac{1}{2} \hbar \omega_{p} 
\;\implies\; 
\int^{\mu} d^3\mathbf{p}\,\rho(p^2)\,\frac{1}{2} \hbar \omega_{p}
\label{MomentumMeasure}
\end{equation}
where the measure factor $\rho(p^2) \sim p^4/M_P^4$, and $\mu \sim M_\mathrm{SM}$.
This measure in momentum space would on one hand imply a curved 
(and thus possibly dynamical) energy-momentum space \cite{g},
and on the other hand the effective freezing of degrees
of freedom with increasing energy, so that the increasing of the 
Planck scale would lower the value of the vacuum energy. 
This counterintuitive behavior is similar to the jamming behavior in non-equilibrium 
statistical physics, in which systems are found to freeze with increasing temperature \cite{jam}. 
It has been argued that such ``freezing by heating'' could be characteristic of a background
independent quantum theory of gravity \cite{chia}.

To elaborate, we note that the usual space-time curvature is defined by the commutator of 
space-time covariant derivatives, $[\,\nabla_{a}, \nabla_{b}\,] \sim R_{ab}$,
and thus the energy-momentum space curvature is expected to come from the
commutator of energy-momentum covariant derivatives.
Given that momentum derivatives usually realize the position operator,
we begin by examining algebraic structures in which $[\,\hat{x}_a, \hat{x}_b\,] \neq 0$.
This in turn implies that the commutation relation between $\hat{x}_i$ and $\hat{p}_j$ 
is different from the canonical \cite{kempf, laynam}.
Let us postulate
\begin{equation}
\dfrac{1}{i\hbar}[\,\hat{x}_i,\,\hat{p}_j\,]\;=\; A(\hat{p}^2)\,\delta_{ij}+B(\hat{p}^2)\,\hat{p}_i\hat{p}_j\;.
\end{equation}
The right-hand side is the most general form depending only on the momentum and that respects
rotational symmetry.  In our previous work, we had 
\begin{equation}
A(p^2)\;=\;1+\beta p^2\;,\qquad B(p^2)\;=\;\beta'
\label{AandB}
\end{equation}
with $\beta$ and $\beta'$ constant,
which introduced a mimimal length $\hbar\sqrt{\beta}$ and
suppressed the vacuum energy at short distances \cite{laynam}.
Here, we allow for a more general form of $A(p^2)$.
Assuming that the momenta commute among themselves,
$[\,\hat{p}_i,\,\hat{p}_j\,] = 0$,
the Jacobi identity tells us that
\begin{equation}
\dfrac{1}{i\hbar}[\,\hat{x}_i,\,\hat{x}_j\,]
\;=\; -\left\{\,2(A+B\hat{p}^2)\dfrac{dA}{dp^2} - AB\,\right\} \hat{L}_{ij}\;,
\end{equation}
where
$\hat{L}_{ij} = \left(\hat{x}_i\hat{p}_j-\hat{x}_j\hat{p}_i\right)/A$
is the generator of rotations.
Following \cite{laynam} we may show that 
the time-evolution invariant phase-space volume in $D$-space dimensions is
\begin{equation}
\dfrac{d^D\mathbf{x}\,d^D\mathbf{p}}{A^{D-1}(A+Bp^2)}\;.
\end{equation}
Setting $B(p^2)=0$ for simplicity,
the vacuum energy density for a massless field will be 
\begin{equation}
\int^{\mu} \dfrac{d^3\mathbf{p}}{A^3(p^2)} \,p 
\;=\; 4\pi \int^{\mu} dp\,\dfrac{p^3}{A^3(p^2)} \;.
\end{equation}
The choice $A(p^2)=1+\beta p^2$ will render this expression finite,
but by itself will not provide
a solution of the cosmological constant problem.
Note that $\rho (p^2) = A^{-3}(p^2)$, \textit{cf.} Eq.~(\ref{MomentumMeasure}), but even
if $A^{-3}(p^2) \sim p^4/M_P^4$, which formally
gives the required cosmological constant seesaw formula
(provided $\mu \sim M_\mathrm{SM}$),
the effective Planck constant $\hbar(p^2)=\hbar\,A(p^2)$ 
blows up at very low momenta, which
clearly contradicts observation.
Thus the theoretical underpinnings of Eq.~(\ref{ccseesaw})
cannot be just based on the minimal length commutators, which
lead to a curved but {\it static} momentum space.
Some other structure is needed and below we argue
that the seesaw formula can be justified in a
{\it dynamical} energy-momentum space.

\begin{figure}[t]
\includegraphics[width=8cm]{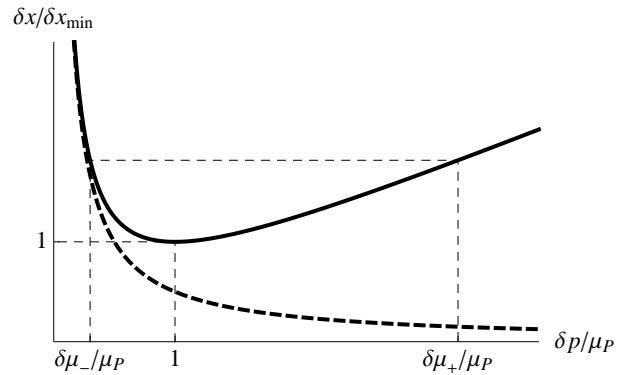}
\caption{The double valued nature of the momentum uncertainty $\delta p$
when one assumes $\delta x\,\delta p = (\hbar/2)(1+\beta\, \delta p^2)$.
The dashed line shows the regular single-valued behavior when $\beta=0$.
Here, $\mu_P\;=\;1/\sqrt{\beta}$ is the Planck momentum, and 
$\delta x_\mathrm{min} = \hbar\sqrt{\beta} = \hbar/\mu_P$.
We argue that the momentum scales corresponding to $\delta\mu_\pm$ are
$(\mu_\pm/\mu_P)\sim(\delta\mu_\pm/\mu_P)^2$.
}
\label{MLUR-fig}
\end{figure}

The minimal length deformation of the canonical commutation
relation leads to a non-canonical {\it double-valued} relation between
the uncertainties for the position and momentum.  
This is graphed in Fig.~\ref{MLUR-fig} for the 
case studied in \cite{laynam}, Eq.~(\ref{AandB}), where
\begin{equation}
\delta x\,\delta p \;=\; \dfrac{\hbar}{2}\left(1 + \beta\, {\delta p}^2\right)\;.
\label{MLUR}
\end{equation}
In particular, for large (i.e. {\it trans-Planckian}) momentum uncertainty, 
$\delta p \gg \mu_P = \beta^{-1/2}$, we have
$\delta x \sim \beta\,\delta p$, 
which encapsulates another feature of string
theory in certain backgrounds \cite{joe}. 
The quantum properties of space-time geometry may be 
understood in terms of effective
expressions that involve the space-time uncertainties:
\begin{equation}
g_{ab}(x)\, d x^a d x^b  \;\to\; g_{ab}(x)\, \delta x^a \delta x^b\;.
\end{equation}
This can now be directly transferred to a {\it local} dynamical geometry of 
energy-momentum space
\begin{equation}
g_{ab}(x)\, \delta x^a \delta x^b \;\to\; G_{ab}(p)\, \delta p^a \delta p^b\;.
\end{equation}
The usual intuition that local properties in space-time
correspond to non-local features of energy-momentum space
(as implied by the canonical uncertainty relations) is
obviated by this linear relation between the uncertainties in coordinate and
momentum spaces.

A dynamical geometry of energy-momentum space suggests a new viewpoint on the cosmological
constant problem that provides quite naturally the seesaw formula in
$3+1$ dimensions and is consistent with relativistic invariance, 
but crucially requires a curved energy-momentum space. 
When quantum effects are turned on, one expects fluctuations of such
a geometry and thus, in general, a foamy energy-momentum space.
Of course, the idea of space-time foam is
quite old \cite{wheeler} and is naturally expected in
a quantum theory of gravity. What we are claiming
is that an analogous ``foaminess'' extends to the energy-momentum
space and is crucial for understanding the problem of vacuum energy.

If space-time is foamy due to quantum
fluctuations \cite{ng}, the fluctuations $\delta l$ will show up when we measure a distance
$l$, in the form of uncertainties in the measurement.
In other words, if one regards the
elementary events partitioning the space-time volume into ``cells,'' then the number of cells is
bounded by the surface area of the spatial region (corresponding to the holographic
scaling $(l/l_P)^2$ of black-hole physics), 
and each cell occupies a space-time
volume of $(l^4/c) /(l/l_P)^2 = l^2 l_P^2/c$ on average.  
The maximum space-time resolution of the geometry is obtained if each clock ticks only once 
during the entire time period $l/c$. 
Then on average each cell occupies a space-time volume no less than 
$
l^4 / (l / l_P)^2 = l^2 l_P^2 \sim \delta l^4
$
yielding an average separation between neighboring cells no less than $l^{1/2} l_P^{1/2}$.  
This spatial separation is interpreted as the average minimum fluctuation $\delta l$ of a distance $l$.
One of the important points of \cite{ng} is that in the case where space and time are
treated on the same footing (that is, relativistically) the scaling of length in the simple holographic models of space-time foam is ``Brownian,''
$
\delta l  \sim l^{1/2} l_P^{1/2}.
$
The crucial fact to note here is that $\delta l$ can be much larger
than the natural Planck length scale $l_p$.
Also, $l$ determines the observable size of space-time.

Given the linear relation between spatial and momentum fluctuations as 
implied by Eq.~(\ref{MLUR}) in the trans-Planckian region, $\delta x \sim \beta\,\delta p$,
we can extend the foaminess in space-time to energy-momentum space.
Thus we have a fully relativistically invariant $3+1$ dimensional
energy-momentum foam characterized by
\begin{equation}
\delta \mu \sim \mu^{1/2} \mu_P^{1/2}.
\end{equation}
From here it follows that the size of energy-momentum space,
i.e. the energy-momentum cut-off scale $\mu$, is given by a seesaw-like formula
$\mu \sim \delta \mu^2/\mu_P$.
Note that in complete analogy with the case of space-time foam \cite{ng}
the fluctuation of the energy-momentum foam $\delta \mu$ is much larger than $\mu_P$.
Thus generically, the energy-momentum fluctuations are indeed trans-Planckian.
However, due to the double-valued nature of Eq.~(\ref{MLUR}),
we have one very large (i.e. trans-Planckian) $\delta \mu_{+}$
and one very small (i.e. sub-Planckian) $\delta \mu_{-}$ such that
$\delta\mu_+\delta\mu_- = \mu_P^2 = \beta^{-1}$.
See Fig.~\ref{MLUR-fig}.
Both of these values satisfy the fundamental relation of the energy-momentum foam
$\delta \mu_{\pm} \sim \mu_{\pm}^{1/2} \mu_P^{1/2}$
where $\mu_{-}$ and $\mu_{+}$ respectively represent
the sub-Planckian and the trans-Planckian momentum cut-offs. 
As we can see, there is a {\it seesaw} relation between
the sub-Planckian (IR) and the trans-Planckian (UV) momentum fluctuations and cut-offs:
\begin{equation}
\mu_{-}\delta \mu_{+}^2 \;\sim\; \mu_{+}\delta \mu_{-}^2 \;\sim\; \mu_P^3\;.
\label{museesaw}
\end{equation}
This relation is absolutely crucial, because if we take the lower value $\delta \mu_{-}$ 
(as we must, because the observed energy-momentum region is sub-Planckian)
we are naturally led to a realistic seesaw formula
for the energy-momentum cut-off $\mu_{-} \sim \delta \mu_{-}^2/\mu_P$,
and the vacuum energy $\epsilon^4 \sim \mu_{-}^4$.
In particular, if we suppose that $\delta \mu_{-}$ is exponentially suppressed from
the Planck scale (perhaps because of its relation to the hierarchy problem) 
so that it is of the order of 
$\delta \mu_{-} \sim M_\mathrm{SM} \sim 1\,\mathrm{TeV}$, then we have a natural solution of
the vacuum energy puzzle, because the vacuum energy is determined as
\begin{equation}
{\epsilon}^4 \;\sim\; \mu_{-}^4 \sim \delta \mu_{-}^8/\mu_P^4\;.
\end{equation}
The vacuum energy naturally vanishes if the Planck scale $\mu_P$ is taken to infinity.
In that case, one expects an extended symmetry that protects the zero value of the vacuum energy.
This symmetry should be an essential feature of a non-perturbative formulation of quantum gravity \cite{chia}.

The above see-saw formula, Eq.~(\ref{museesaw}), could be understood physically
by appealing to the linear relation between the space-time and
energy-momentum uncertainties in the trans-Planckian region:
$\delta x \sim \beta\,\delta p$. In that case, the growing uncertainty
in position modes is related to the growing uncertainty in the momentum modes. 
This stretching of the trans-Planckian modes effectively causes the `jamming,' 
or `freezing,' of the low energy sub-Planckian modes so that the effective low-energy
momentum shell is given by a naturally small $\mu_{-}$.
Note that the trans-Planckian shell is invisible from the point of
view of the low energy modes.
In some sense the $\mu_{-}$ is the inner, and $\mu_{+}$ the outer
horizon in the dynamical energy-momentum space.


A dynamical energy-momentum space will have significant phenomenological consequences.  
Among these would be how Eq.~(\ref{MLUR}) could be accommodated in quantum field theory. 
The usual relation $\delta x\,\delta p = \hbar/2$,
shown with a dashed line in Fig.~\ref{MLUR-fig}, is a simple consequence of the fact that
coordinate and momentum spaces are Fourier transforms of each other.  The more one wishes to 
localize a wave-packet in coordinate space, the more momentum states one must superimpose.
In the usual case, there is no lower bound to $\delta x$: one may localize the wave-packet as
much as one likes by simply superimposing states with ever larger momentum, and thus ever shorter wavelength, to cancel out the tails of the coordinate space distributions.  On the other hand, 
the deformed uncertainty relation Eq.~(\ref{MLUR}), shown with a solid line in Fig.~\ref{MLUR-fig}, 
implies that if one keeps on superimposing states with momenta beyond the Planck momentum 
$\mu_P = 1/\sqrt{\beta}$,
the uncertainty in position $\delta x$ will cease to decrease and start increasing instead.
The natural interpretation of such a behavior would be that the trans-Planckian modes ($p>\mu_P$) when superimposed with the sub-Planckian ones ($p<\mu_P$)
will `jam' the sub-Planckian modes and prevent them from canceling out the tails of the wave-packets effectively.

\begin{figure}[t]
\includegraphics[scale=0.9]{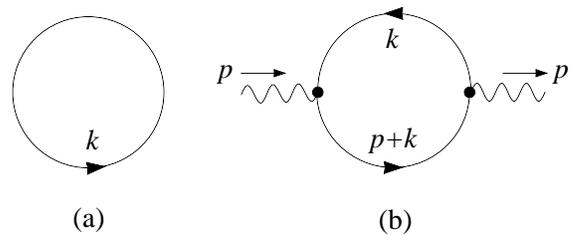}
\caption{(a) Vacuum fluctuation diagram, and (b) vacuum polarization diagram with external legs. 
If the contribution of modes beyond $\mu_-$ is somehow suppressed in (a), a similar suppression
should be expected in (b).}
\label{loops}
\end{figure}

This `jamming' mechanism explains why the effective UV cut-off for the
calculation of the cosmological constant is the sub-Planckian $\mu_-$ and not
the trans-Planckian $\mu_+ \sim \mu_P^2/\mu_-$. 
The cosmological constant can be thought of as the contribution of
the vacuum fluctuation diagram shown in Fig.~\ref{loops}(a) to the effective potential. 
Since the loop momentum $k$ can take on any value, both sub-Planckian and trans-Planckian
momentum modes can be expected to contribute.  However, due to `jamming,' the contributions
of the sub-Planckian ($\mu_-<k<\mu_P$) and trans-Plankian ($\mu_P<k<\mu_+$) 
modes cannot be simply added together as
those due to independent degrees of freedom.  Rather, they can be expected to
`jam' each other, and their contributions to cancel out, effectively lowering the
UV cut-off from $\mu_+$ to $\mu_-$.  
While we have proposed a specific realization of this behavior through a minimal length deformation, the fact that $\Lambda$ 
is as small as it is implies that such a cancellation
must take place in any dynamical description involving gravity.  

The lowering of the effective UV cut-off in Fig.~\ref{loops}(a) due to `jamming,'
or via any other mechanism for that matter, 
should also occur in diagrams with external legs, like those 
shown in Fig.~\ref{loops}(b), the only difference being the external momentum flowing through the diagram.  
Unlike Fig.~\ref{loops}(a), however,
where the cancellation of the sub- and trans-Planckian contributions is almost perfect with the 
cut-off $\mu_-$ in the meV range, the external momentum must destroy the near-perfect cancellation 
and raise the effective cut-off to a much higher scale since the running of couplings
due to diagrams such as Fig.~\ref{loops}(b) has already been observed.

\begin{figure}[t]
\includegraphics[scale=0.9]{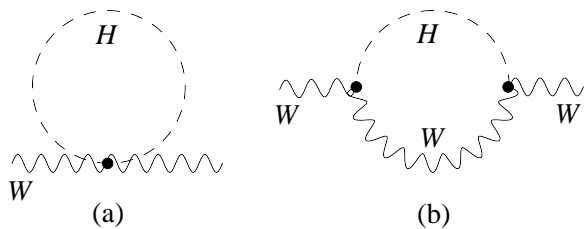}
\caption{Higgs contribution to the $W$ propagator. In non-unitary gauges, there is another diagram
in which the $W$ propagator in the loop in (b) is replaced by that of a goldstone.}
\label{higgs}
\end{figure}

Nevertheless, given that the effective cut-off for Fig.~\ref{loops}(a) is so small,
it would be quite mysterious if the effective cut-offs for diagrams of the sort in Fig.~\ref{loops}(b) 
were much higher than the electroweak scale. 
Indeed, a hint of a low effective UV cut-off may have already been observed:
it is well known that the preferred mass of the Standard Model Higgs boson obtained from a global fit to all precision electroweak data is lower than the direct search bound established by 
LEP2 and the Tevatron \cite{RPP}.
Recalling that the dependence of the SM predictions on the Higgs mass is logarithmic and enters 
via oblique correction diagrams such as those shown in Fig.~\ref{higgs}, 
one can intepret the Higgs mass itself
as an effective UV cut-off rendering the SM predictions finite.  
Thus the low Higgs mass prefered by the SM fit could be a manifestation of a truely low effective UV cut-off.

It is clear that the proposed mechanism poses an alternative solution to the hierachy problem without imposing supersymmetry, although the scheme
could of course accommodate the addtional symmetry.  Details on this how this comes about, as well as phenomenological implications in other contexts, 
such as the Casimir effect, will be addressed elsewhere.


\medskip
\noindent
\textbf{Acknowledgments:}
We wish to thank Sumit Das, Martin Kruszenski, Oleg Lunin, Samir Mathur, Jack Ng, Al Shapere
and Chia Tze for comments regarding this work.
DM and TT are supported in part by
the U.S. Department of Energy, grant DE-FG05-92ER40677, task A.



\begin{thebibliography}{99}
\bibitem{planck}
There is an intriguing similarity here with Planck's resolution of
the UV catastrophe of the black body radiation.
\end{thebibliography}
\end{document}